\documentclass[aps,prl,twocolumn,groupedaddress]{revtex4}

\usepackage{graphicx}
\usepackage{epstopdf}
\usepackage{color}
\usepackage{natbib}
\usepackage{hyperref}
\usepackage{soul}
\usepackage{float}

\begin{document}

%\preprint{}

\title{Counterintuitive temporal shape of single photons}

\author{Gurpreet Kaur Gulati}
\altaffiliation{Center for Quantum Technologies 3 Science Drive 2, Singapore,
  117543}
\author{Bharath Srivathsan}
\altaffiliation{Center for Quantum Technologies 3 Science Drive 2, Singapore,
  117543}
\author{Brenda Chng}
\altaffiliation{Center for Quantum Technologies 3 Science Drive 2, Singapore,
  117543}
\author{Alessandro~Cer\`e}
\altaffiliation{Center for Quantum Technologies 3 Science Drive 2, Singapore,
  117543}
\author{Dzmitry Matsukevich}
\altaffiliation{Department of Physics, National University of Singapore, 2
  Science Drive 3,  Singapore, 117542}
\altaffiliation{Center for Quantum Technologies, National University of
  Singapore, 3 Science Drive 2, Singapore, 117543}
\author{Christian Kurtsiefer}
\altaffiliation{Department of Physics, National University of Singapore, 2
  Science Drive 3,  Singapore, 117542}
\altaffiliation{Center for Quantum Technologies, National University of
  Singapore, 3 Science Drive 2, Singapore, 117543}
%\thanks{}
%\email[]{christian.kurtsiefer@gmail.com}
\date{\today}

\begin{abstract}
We prepare heralded single photons from a photon pair source based on
non-degenerate four-wave mixing in a cold atomic ensemble via a cascade decay
scheme. Their statistics shows strong anti-bunching with $g^{(2)}(0)<0.03$,
indicating a near single photon character. In an optical homodyne experiment,
we directly measure the temporal envelope of these photons and find,
depending on the heralding scheme, an exponentially decaying or rising
profile. The rising envelope will be useful for efficient interaction between
single photons and microscopic systems like single atoms and molecules. At the
same time, their observation  illustrates the breakdown of a realistic
interpretation of the heralding process in terms of defining an initial
condition of a physical system.

\end{abstract}


\maketitle

Strong atom-light interaction at the single quantum level is a prerequisite for
many quantum communication 
and computation protocols~\cite{Cirac_1997,DLCZ:2001,Wilk_2007,kimble2008}.
In free space the optimal coupling of photons to atoms requires 
the temporal profile of the incoming photons to match the time reversal of photons generated by spontaneous decay from the transition of interest.~\cite{Yimin:2011,Syed_2013,Bader_2013}.
Together with spectral and spatial mode matching~\cite{Nicolas_2001,Sondermann:07,Mengkhoon_2008}, the efficient preparation of single photons with an exponentially rising temporal envelope 
remains an open challenge. 

A common way to obtain single photons is to generate time correlated photon
pairs: the detection of one photon heralds the presence of the
other~\cite{Clauser_1974,Grangier_1986,Hong:86}.
Single photons with an exponentially rising temporal envelope have been prepared from such a heralded single photon source by direct modulation of one of the photons of the pair~\cite{Kolchin_2008}. In the work presented here, we use a photon pair source based on four-wave mixing in a cold atomic ensemble via a cascade decay level scheme.
This process has been used in the past to generate narrow band photon
pairs~\cite{Chaneliere:2006}, and it has already been demonstrated that 
the resulting photon pairs are nearly Fourier-limited~\cite{Srivathsan:2013}
with a coherence time long enough to be resolved with various optical
detection techniques. We demonstrate how this source allows the direct preparation of single photons with a rising exponential temporal envelope.

\begin{figure}[]
  \begin{center}
    \includegraphics[width=\columnwidth]{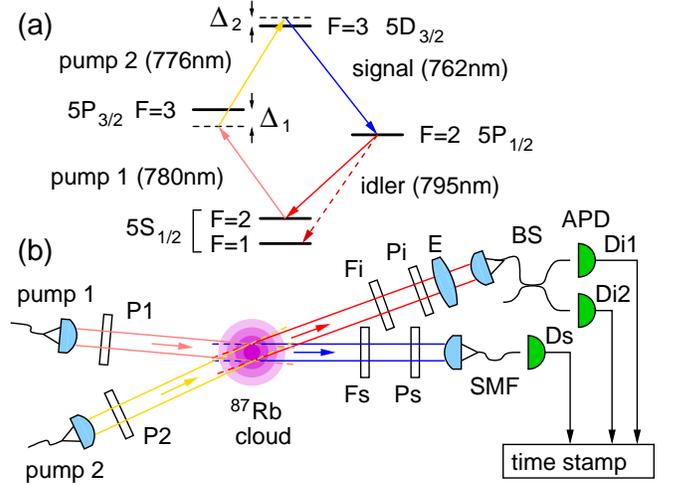}
   \caption{\label{fig:level_scheme_HBT}(a) Cascade level scheme for four
     wave mixing in$^{87}${Rb}. (b) Setup of the heralded single photon
     source. Polarizers (P1,2, Pi, Ps) and interference filters (Fi, Fs)
     separate the signal and idler photons from residual pump light. 
     An etalon (E) in the idler arm removes uncorrelated photons from a decay to
     the $5\text{S}_ {1/2},\,F=1$ level. Signal photons detected by an
     avalanche photodiode (APD) trigger a Hanbury-Brown--Twiss measurement
     between idler photons distributed to two detectors $D_{\text{i1}},
     D_{\text{i2}}$ via a 50:50 fiber beam splitter (BS).}
 \end{center}
\end{figure}

We initially demonstrate (see Fig.~\ref{fig:level_scheme_HBT}) the single photon character of our heralded photons. 
An ensemble of $^{87}\text{Rb}$ atoms is
first cooled and confined in a magneto-optical trap (MOT) reaching an
optical density of $\approx$ 32 on the $5\text{S}_{1/2},\,F=2\rightarrow
5\text{P}_{3/2},\,F=3$ transition after about 12\,ms.
In the following 1\,ms the MOT beams are switched off, and the atoms are excited to the $5\text{D}_{3/2},F=3$ level by two orthogonally linearly
polarized pump beams (780\,nm, 0.1\,mW and 776\,nm, 5\,mW, beam waists 0.45\,mm) intersecting
at an angle of 0.5$^\circ$ in the cold atomic cloud. 
The 780\,nm pump beam is red detuned by $\Delta_1=40$\,MHz from
the intermediate level $5\text{P}_{3/2},\,F=3$ to reduce
incoherent scattering. The combined detuning of both
pumps from the two-photon transition ranges from $\Delta_2=0-6$\,MHz to the blue.
Photon pairs of wavelength 762\,nm (signal) and 795\,nm (idler) are generated by a cascade decay from the $5\text{D}_{3/2},F=3$ level via $5\text{P}_{1/2},F=2$ to $5\text{S}_{1/2},F=2$. 
We use interference filters and an etalon to filter any background light generated by other processes in the ensemble. 
Energy conservation and phase matching between pump and collection modes allow efficient coupling of photon pairs with a strong temporal correlation into
single mode fibers. 

Unlike single quantum emitters
\cite{Kurtsiefer_2000,Michler_2000,Kuhn_2002}, the probability of generating
more than one photon per heralding event in a parametric process does not
vanish due to the thermal nature of the emission process from the atomic
ensemble \cite{mandel_wolf}. We consider the second order
correlation function $g^{(2)}(\Delta t_{12})$ for the probability of
observing two photons in a given mode with a time difference $\Delta t_{12}$.
Any classical light field exhibits $g^{(2)}(0)\ge1$, while
$g^{(2)}(\Delta t_{12})<1$ is referred to as photon antibunching, with an ideal
single photon source reaching $g^{(2)}(0)=0$ \cite{Glauber_1963}. 

We determine this correlation function experimentally in a
Hanbury-Brown--Twiss (HBT) geometry, where the idler light is distributed
with a 50:50 fiber beam splitter onto two single photon counting silicon
avalanche detectors (APD) $D_{\text{i1}}, D_{\text{i2}}$ (quantum efficiency
$\approx\,\,$40\%, dark count rates 40 to 150\,s$^{-1}$),  while signal photons are
detected by $D_{\text{s}}$ as heralds. The detection events are
time-stamped with 125\,ps resolution. The combined timing uncertainty
of the photodetection process is $\approx$ 600\,ps.

From our previous characterization of the source~\cite{Srivathsan:2013}, 
we know that
the correlation function between the signal and idler $g^{(2)}_{si}(\Delta
t_{si})$ has the shape of a decreasing exponential, with more than 98\% of the coincidences occuring within a time window $T_c$=30\,ns. 
We record a histogram $G^{(2)}_{\text{i1i2$|$s}}(\Delta t_{12})$ of
idler detection events on $D_{\text{i1}}$ and $D_{\text{i2}}$ with
a time difference $\Delta t_{12}=t_2-t_1$ if
one of them occurs within a coincidence time window $T_c$
after the detection of a heralding event in the signal mode. 
The normalized correlation function of heralded coincidences between
the two idler modes is
\begin{equation}
g^{(2)}_{\text{i1i2$|$s}}(\Delta t_{12}) =G^{(2)}_{\text{i1i2$|$s}}(\Delta
t_{12}) / N_{i1i2|s}(\Delta t_{12}),
\end{equation}
where $N_{i1i2|s}(\Delta t_{12})$ is the estimated number of
accidental coincidences. 
Due to the strong temporal correlation between signal and idler photons, the probability of accidental coincidences is not uniform.
We thus estimate $N_{i1i2|s}(\Delta t_{12})$ for every $\Delta t_{12}$ by
integrating the time difference histograms between the signal and each arm of
the HBT, $G^{(2)}_{si1}\left(\Delta t_{si}\right)$ and
$G^{(2)}_{si2}\left(\Delta t_{si}\right)$ within $T_c$.

Due to the time ordering of the cascade  process, it is only meaningful to consider positive time delays after the detection of the heralding photon,
thus splitting $N_{i1i2|s}$ into two cases. For $\Delta t_{12}\geq0$, we use
\begin{equation}
  N^{(+)}_{i1i2|s}(\Delta t_{12})=\int\limits_0^{T_c}
  G^{(2)}_{si1}\left(\Delta t_{si}\right)
  G^{(2)}_{si2}\left(\Delta t_{si}+\Delta t_{12}\right)\; d\Delta t_{si}
\end{equation}
while for $\Delta t_{12}<0$, we use
\begin{equation}
  N^{(-)}_{i1i2|s}(\Delta t_{12})=\int\limits_0^{T_c}
  G^{(2)}_{si1}\left(\Delta t_{si}+\Delta t_{12}\right)
  G^{(2)}_{si2}\left(\Delta t_{si}\right)\; d \Delta t_{si}.
\end{equation}

\begin{figure}
  \begin{center}
    \includegraphics[width=\columnwidth]{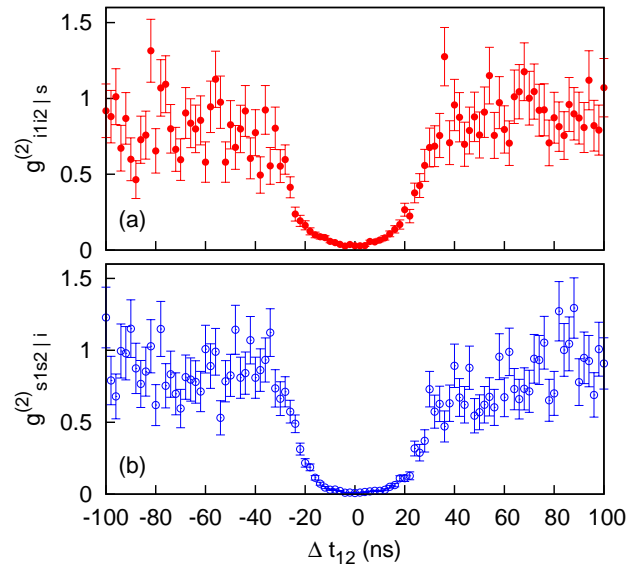}
    \caption{\label{fig:conditional_g2_762} (a) The correlation function
      $g_{i1i2|s}^{(2)} $ of idler photons separated by a time difference
      $\Delta t_{12}$, conditioned on detection of a heralding event in the
      signal mode, shows strong photon antibunching  over a time scale of
      $\pm20$\,ns, 
      indicating the single photon character of the heralded photons.
	The errorbars indicate the propagated poissonian counting uncertainity from $G^{(2)}_{i1i2|s}$ and $N_{i1i2|s}$.    
	 (b) Same measurement, but with the signal and idler modes swapped.}
  \end{center}
\end{figure}

The resulting $g^{(2)}_{\text{i1i2$|$s}}(\Delta t_{12})$ is shown in
Figure~\ref{fig:conditional_g2_762}(a) as function of the
delay $\Delta t_{12}$, sampled into 2\,ns wide time bins. 
With a signal photon detection rate of 50000\,s\,$^{-1}$ (at $\Delta_2=0$), we
observe $g^{(2)}_{\text{i1i2$|$s}}(0)=0.032\pm0.004$. 

When switching the roles of the signal and idler arms, the resulting
normalized correlation function shown in
Figure~\ref{fig:conditional_g2_762}(b) has a minimum $g^{(2)}_{\text{s1s2$|$i}}$ of
0.018$\pm$0.007 with an idler photon detection rate of
13000\,s$^{-1}$.

\begin{figure}[]
  \begin{center}
    \includegraphics[width=0.8\columnwidth]{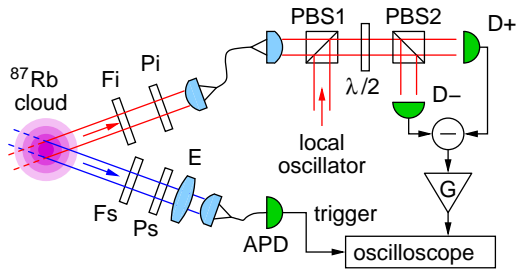}
   \caption{\label{fig:homodyne_setup}
     Field measurement setup. One of the photons (idler in this figure) is
     combined with a coherent laser field as a local oscillator on a polarizing beam splitter (PBS1), and sent to a balanced pair of pin photodiodes D+, D- for a homodyne measurement. The
     photocurrent difference as a measure of the optical field strength is
     amplified and recorded with an oscilloscope on the arrival of a heralding
     event from the avalanche photodetector in the signal arm.}
 \end{center}
\end{figure}

\begin{figure}
  \begin{center}
	\includegraphics[width=\columnwidth]{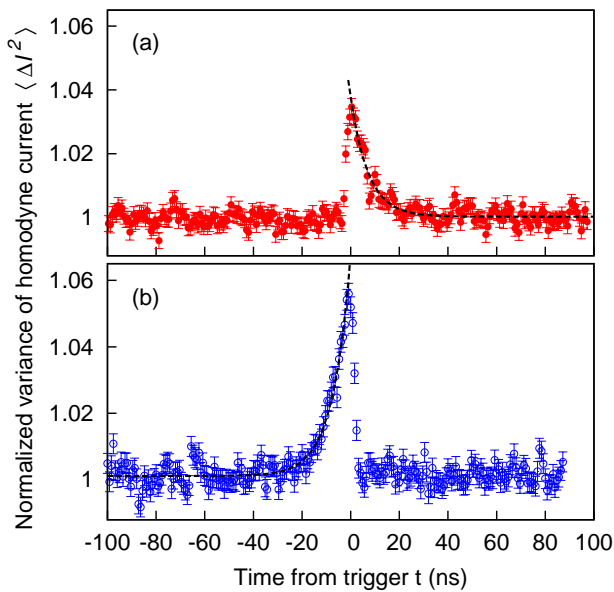}
   \caption{\label{fig:combined_field}Optical homodyne results. (a)
     Exponential decay of the field variance of heralded idler photons. (b)
     Exponential rise of the field variance of heralded signal photons.}   
  \end{center}
\end{figure}

We now proceed to determine the temporal envelope of the heralded single
photon fields by measuring the field quadrature in the time domain
\cite{Yuen:83,lvovsky:2001} via a balanced homodyne detection.
The experimental scheme is shown in Figure~\ref{fig:homodyne_setup}: the
idler mode is mixed with a local oscillator (LO)
which is frequency-stabilized to the idler transition
$5\text{S}_{1/2},\,F=2$ $\rightarrow$  $5\text{P}_{1/2},\,F=2$
The balanced mixing is done with two polarizing beam splitters (PBS1,2)
and a half-wave plate with an interference visibility of
$\approx$~95\%.
The difference of photocurrents from pin silicon photodiodes (D+, D-;
quantum efficiency $\approx$ 87\%) is proportional to the optical field
quadrature in the idler mode.
 With a LO power of 4.5\,mW, the electronic noise is about 6-20dB below the
 shot noise limit over a band of 10\,kHz--210\,MHz.
We record the homodyne signal with a digital oscilloscope (analog
bandwidth 1\,GHz), with the click detection of the signal photon on the APD
triggering the acquisition.
We then calculate the variance of the optical field from $2.7\times10^5$ traces,
normalized to the shot noise, as a measure of the
temporal envelope of the photon. We also switched the
roles of the signal and idler modes for triggering and homodyne detection, this time using a local oscillator resonant with the  transition $5\text{P}_{1/2},\,F=2$$\rightarrow$
$5\text{D}_{3/2},\,F=3$ near 762\,nm.  The variance for this measurement is calculated from $5\times10^5$ traces.
Both results are shown in Figure
~\ref{fig:combined_field}. In both configurations, we set
$\Delta_2\approx6$\,MHz to maximise the heralding efficiency.

In Figure~\ref{fig:combined_field}(a) the normalized field variance of the idler photon suddenly rises about 4\% above the shot noise level at the detection time of the trigger photon, and
exponentially decays back to the shot noise level, with a time constant of
$\tau_i=7.2\pm0.2$\,ns obtained from a fit. This can be easily understood by
the timing sequence of a cascade decay, where the signal photon heralds the
population of the intermediate level, which subsequently decays exponentially,
leading to the characteristic decaying envelope of the idler photon according
to the Weisskopf-Wigner solution \cite{weisskopf:30,Franson_1989}. 
The time constants for the exponential rise or fall times are compatible with the
distribution of detection time differences we observe in the cascade decay
\cite{Srivathsan:2013}, enhanced by the dense atomic ensemble \cite{Dicke:53}.
The probability of detecting an idler photon given the detection of a signal
photon (heralding efficiency) was independently determined with two APDs
 to be $\eta_{i}\approx13\% $ (uncorrected for APD efficiency and optical losses).

In Figure~\ref{fig:combined_field}(b) the normalized field variance of the signal photon
exponentially increases a few 10\,ns before the trigger event to a value of
1.06, and then quickly returns to the shot noise level, with a rise time constant
$\tau_s=7.4\pm0.2$\,ns obtained from a fit. Here, the suppression of
uncorrelated trigger (idler) photons by the etalon E (figure
\ref{fig:level_scheme_HBT}(b)) results in a higher heralding efficiency
$\eta_{s}\approx19\%$, and therefore a higher signal to shot noise level. 

From the results of the HBT experiment, we know that the idler detection witnessed a single photon to a very good approximation. We therefore have to conclude that the heralded signal field is a single photon state with an exponentially rising temporal envelope as required for optimal absorption with single atoms or molecules.

In this case, however, the simple causal interpretation of the physical
process in the Weisskopf-Wigner picture does not work: The trigger
time is fixed by the herald that leaves the atoms in the ground state, but the
signal field  starts to rise to a maximum before that. So the
heralding process does not set an initial condition of a physical system that
then evolves forward in time, but marks the end of a (signal) field evolution
that is compatible with the exponential rise that started
{\em before} the heralding event. Formally this is not a problem, because the
heralding event just sets a different boundary condition.

This experiment again highlights a problem with the
definition of  ``real'' physical quantities (in the spirit of an EPR definition
\cite{einstein:35}). The physical quantity here is the electrical
field in the signal mode at any point in time. Nothing seems to set the initial
condition leading to such an increase, with a dynamics governed by
some laws of physics. Yet, when an idler event is registered in a
photodetector, the recorded field is perfectly compatible with a single photon
with an exponentially rising envelope. 
In this example, an interpretation
that is more symmetric between preparation and detection
procedures~\cite{Watanabe:55}, like the two-state vector
formalism~\cite{Reznik:95}, may be adequate.

In summary, we have demonstrated a source of heralded single
photons based on an ensemble of cold rubidium atoms. We observe antibunching
with $g^{(2)}(0)<0.03$, conditioned on the
photon in the signal (idler) mode.
Depending on which of the modes is chosen as a herald, we find either an
exponentially decaying or rising temporal envelope of the heralded photon. If
heralded single photons are practically not distinguishable from ``true''
single photons, the latter should --- at least in principle --- efficiently
be absorbed by a single atom in free-space in a
time-reversed Weisskopf-Wigner situation. Such an experiment would therefore
not only demonstrate strong atom-photon interactions, but
also provide a better understanding to what extent heralded photons are
equivalent to single photons emerging from a setup with a well-defined initial
condition.

We acknowledge the support of this work by the National Research Foundation \&
Ministry of Education in Singapore.
\bibliographystyle{apsrev}
% \bibliography{homodyne}{}

\end{document}